\documentclass{article}
\usepackage{amsmath}
\usepackage{amsfonts}
\begin{document}
\newcommand{\p}{\partial}

\title{\LARGE \bf Non-Abelian Monopole Equations\\ with Zero
Curvature and \\ Self-Dual Yang-Mills Theories}

\author{M. Legar\'e\\Mathematical and Statistical Sciences\\U. of
Alberta, Edmonton\\Canada, T6G 2G1}
\date{November 2001}

\maketitle
\begin{abstract}
A version of non-Abelian monopole equations is explored
through dimensional reductions, with often the addition of
algebraic conditions. On zero curvature spaces, spinor related
extensions of integrable systems have been generated, and
certain reduced one-dimensional systems have been discussed
with respect to integrability, as well as solutions found.
\end{abstract}

\section{Introduction}

The Abelian monopole, or Seiberg-Witten, equations have been
found in relation with Donalson-Witten theories, which
themselves can be interpreted as twisted $N=2$ supersymmetric
Yang-Mills systems (see for instance refs \cite{W1,Do,BBRT,W2}).
These monopole equations can be formulated through a topological
action \cite{LM1,Ma1} and, as for the Donalson-Witten theories,
can be useful in describing invariants as well as aspects of the
topology of four manifolds \cite{W1,Do,Mo}.

Versions of extensions of the Seiberg-Witten equations to
non-Abelian mono\-pole equations have been obtained (see refs
\cite{LM2,BGP,OT}). For the version of refs \cite{Ma1,LM2}, a
topological action based on the Mathai-Quillen formalism has
been shown, and found to be described as twisted $N=2$
supersymmetric gauge theories wih matter hypermultiplet
\cite{Ma1,LM2}. Topological Yang-Mills theories, or
Donaldson-Witten theories, are known to explore the moduli
space of (anti-) self-dual Yang-Mills equations, which are
completely solvable. Moreover, the Abelian and
non-Abelian monopole equations correspond to equations probed
by the (above-mentioned) topological actions, and in the limit
of vanishing spinors, coincide with the (anti-) self-dual
Yang-Mills equations. One can also mention that Abelian
monopole equations have been linked to integrable systems via
preprotentials (see for example, refs \cite{MM} and \cite{HP})

In what follows, the main objective is to present first steps
in exploring non-Abelian monopole equations, for which a
Mathai-Quillen interpretation exists \cite{Ma1,LM2}, with
respect to integrability, partial or complete, as well as
finding solutions.

The method of reductions (e.g. \cite{O1}), here more precisely,
dimensional reductions accompanied by certain sets of algebraic
constraints, of non-Abelian monopole equations with gauge
group $SU(n)$ is used. This approach has also been carried out
to exhibit solutions and show correspondences for the Abelian
monopole equations (\cite{T1,O2,NS,F1,AMN}). For their
integrability behaviour, in addition to the Painlev\'e PDE
test, it is for example of interest to recall that the property
of Painlev\'e has been conjectured to appear in the reductions
of completely integrable systems (Painlev\'e ODE Test)
\cite{AC}, a result which has been proved for certain cases.
Thus, one could for instance imagine checking the reductions of
non-Abelian monopole equations for such property.

\section{Non-Abelian Monopole Equations}

On a four-dimensional oriented, closed manifold $M$, endowed
with a Riemannian metric $g$, a spin connection $\omega$ is
defined. (When no spin structure exists, a spin$^c$ structure is
introduced instead, but only spin manifold are here considered.)
On the tensor product of the Dirac bundle $S = S^+ \oplus S^-$,
with an associated bundle $E$ to a (non-Abelian) structure
group $G$ and base manifold $M$, one provides complete
covariant derivatives, denoted $\mathcal{D}_\mu$, on
multiplets of (commuting) Weyl spinors, $M_\alpha^i \in
\Gamma(S^+\otimes E), i=1,...,N$, transforming under the
$N$-dimensional representation of $G$ :
\begin{equation}\label{complcovder}
{\mathcal D}_\mu M_\alpha^i = \p_\mu M_\alpha^i - \frac{i}{2}
\omega_\mu^{mn} (\sigma_{mn})_\alpha\,^\beta M_\beta^i + i
A_\mu^{ij} M_\alpha^j,
\end{equation}
where the connection $A$ has components $A^{ij}_\mu$ on $E$.
(Notations are detailed in refs \cite{Ma1,LM2}.)

The non-Abelian equations on the spin manifold $M$ can then be
expressed with respect to a Hermitian basis $\{T^a\}, a=1,...,
n^2-1$ of the Lie algebra of $SU(n)$ as :
\begin{equation}
(F_{\alpha\beta}^{+\,a}) + i\bar M_{(\alpha}^i (T^a)^{ij}
M_{\beta)}^j = 0,
\label{nameq1}\end{equation}
\begin{equation}
{\mathcal D}^{\dot\alpha\alpha} M_\alpha = 0,
\label{nameq2}\end{equation}
which involve the self-dual gauge field strength
$F_{\alpha\beta}^+$, and the positive chirality (Weyl) commuting
spinor $M_\alpha$. Also : $\bar M^{\alpha i} = (a^{i\ast},
b^{i\ast}), \bar M_\alpha = \bar M^\beta
(\sigma_2)_{\alpha\beta}$, and $\bar M^i_{(\alpha} M^j_{\beta)}
= \bar M_\alpha^iM_\beta^j + \bar M_\beta^i M^j_\alpha$. The
Dirac operator can be expressed as follows in terms of the
tetrads $e^{m\mu}$ on $M$ and $\Sigma_0 = \sigma_0, \Sigma_i =
i\sigma_i, i=1,2,3$ :
\begin{equation}
{\mathcal
D}^{\dot\alpha\alpha} = (\Sigma_m)^{\dot\alpha\alpha} e^{m\mu}
{\mathcal D}_\mu,
\end{equation}

Let us add that $SU(n)$ gauge transformations will preserve
the set of equations (2,3). There is moreover a corresponding
topological action for the above non-Abelian monopole equations
\cite{Ma1,LM2}, preserved by $Q-$ (or BRST-like) transformations.

Under reductions, sets of $Q-$ transformations, left untouched
by the symmetry or algebraic conditions, will be allowed if
they commute with translations along the coordinates $x^\mu$
\cite{LL}. Rotations can be such symmetries leaving (residual)
$Q-$ transformations. However, the presence of rotations in the
isotropy subgroup could impose vanishing invariant Weyl spinors
($M_\alpha^i$), and therefore the reduced systems probed by the
reduced topological action would correspond to the reduced
(anti-) self-dual Yang-Mills equations.

\section{Example of Reduced Non-Abelian Monopole Equations}

In this example, reductions by translations along the
coordinates $x^1, x^2, x^3$ are performed. The reduced
equations have the following general form with respect to a
Hermitian basis of the Lie algebra of the gauge group $SU(n)$,
and with gauge choice $A_0 = 0$ :
\begin{align}\label{nameq-1}
(D_0 A_1 + i[A_2,A_3])^a &= (a^{i\ast} (T^a)^{ij} b^j  + b^{i\ast}
(T^a)^{ij} a^j),\nonumber\\
(D_0 A_2 + i[A_3,A_1])^a &= -i(a^{i\ast} (T^a)^{ij} b^j  -
b^{i\ast} (T^a)^{ij} a^j),\\
(D_0 A_3 + i[A_1,A_2])^a &= (a^{i\ast} (T^a)^{ij} a^j  -
b^{i\ast} (T^a)^{ij} b^j),\nonumber
\end{align}
for the reduced equations from the (anti-) self-dual Yang-Mills
system with spinor extensions. For the Dirac equation in the
background of the gauge fields, one finds :
\begin{equation}\label{nameq-2}
\left [ \begin{matrix} D_0-A_3 & -A_1+iA_2 \\ -A_1-iA_2 & D_0+A_3
\end{matrix} \right ]^{ij} \left [\begin{matrix} a^j \\ b^j \end{matrix}
\right ]  = 0,
\end{equation}
where the covariant derivative $D_\mu$ involves only the spin
connection. The whole system of equations (5,6) could be seen
(since in 1d) as a ``spinor" extension of the Nahm's equations.

If one chooses in addition to translation invariance, the
conditions $A_2=A_3=0$, and reverts the gauge choice on $A_0$,
to put : $A_0 = -i N$, and $A_1 = L$, with the constraint on
spinor contribution components : $a =\pm b$, then a spinor
extended form of integrable systems related to Lax pairs
$(L,N)$, is derived :
\begin{align}
\left [\dfrac{d}{dx^0} + M, L \right ]^a &= \pm 2 a^\dagger T^a
a,
\\
\left (\dfrac{d}{dx^0} + M \mp L\right )^{ij} a^j &= 0,
\end{align}
where (7) is a Lax pair equation with spinor contributions.
Different examples of reductions can be found in ref. \cite{L1}.

\section{Solution on Flat Space}

In this section, one example of solution to a set of reduced
non-Abelian monopole equations on flat space is shown. The
o.d.e's are obtained by imposing  $SU(2)$ as gauge group, along
with the following gauge fields :
\begin{equation}
A_1 = (-X)\sigma_1, \quad A_2 = (-X) \sigma_2, \quad
A_3 = (-Y) \sigma_3,
\end{equation}
where $X,Y$ are functions of $x^0$ only. The spinor
contributions are given as : $a^1=b^2 =A$, and $a^2=b^1 = B$,
where $A$ and $B$ $\in \mathbb R$.

This system of o.d.e.'s can be interpreted as an extension of
the $SU(2)$ Toda integrable equations, which can be written as :
\begin{align}\label{toda-su2}
d_0 Y + 2X^2 &= 2(B^2-A^2), \nonumber\\
d_0 X + 2XY &= -2(A^2+B^2),\\
d_0 X + 2XY &= 2(A^2-B^2),\nonumber 
\end{align}
deduced from the self-dual gauge fields equations; and for its
reduced Dirac part :
\begin{equation}
(d_0+Y)A = 0 ;\qquad (d_0+2X-Y)B =0.
\end{equation}

Then, from the solution $A=0$, and the supplementary condition
: $X=-Y$, one derives that :
\begin{align}\label{toda-simpl}
d_0 Y + 2Y^2 &= 2B^2,\\
d_0 B - 3Y B &= 0.
\end{align}

The latter equations (12,13) lead to the following o.d.e. for
$Y$ :
\begin{equation}
d_0^2 Y - 2Y d_0Y - 12 Y^3 = 0,
\end{equation}
for which solutions can be found in terms of elliptic
integrals, but which has not been found to be of the Painlev\'e
type. When the spinor contribution $B$ vanishes, the
corresponding equation for $Y$ is of the Painlev\'e type, as
expected from the integrability of the self-dual Yang-Mills
equations.

\section{Conclusion}

More examples of reductions and solutions are presented in ref.
\cite{L1}. Further developments could be for instance in the
direction of non-Abelian monopole equations on curved spaces,
whether or not a spin structure exists (e.g. as generalizations
of compactifications \cite{Ma2}). Their relations to
integrable systems, as extended systems based on spinor
components, and the study of their integrability property could
also be the focus of different activities.

\section{Acknowledgments}

The author acknowledges for the support of this work a grant from 
the National Sciences Research Council (NSERC) of Canada.

\end{document}